\documentclass[sigconf,authorversion]{acmart}
\usepackage{multirow}
\usepackage{todonotes}
\usepackage{subcaption}

\newcommand{\gball}{\,\textcolor{green}{\begin{picture}(-1,1)(-1,-3)\circle*{5}\end{picture}\ }}
\newcommand{\oball}{\,\textcolor{orange}{\begin{picture}(-1,1)(-1,-3)\circle*{5}\end{picture}\ }}
\newcommand{\rball}{\,\textcolor{red}{\begin{picture}(-1,1)(-1,-3)\circle*{5}\end{picture}\ }}
\newcommand{\bball}{\,\textcolor{darkgray}{\begin{picture}(-1,1)(-1,-3)\circle*{5}\end{picture}\ }}

\newcommand{\tableturnl}{-90}

\AtBeginDocument{%
  }

\copyrightyear{2023}
\acmYear{2023}
\acmBooktitle{This is the author's version of the work. It is posted here for your personal use. Not for redistribution. The definitive Version of Record was published at ARES 2023 resp. ACM.}
\acmPrice{15.00}
\acmDOI{10.1145/3600160.3605024}

\begin{document}

\title{Evaluation of Real-World Risk-Based Authentication at Online Services Revisited: Complexity Wins}

\author{Jan-Phillip Makowski}
\author{Daniela Pöhn}
\orcid{0000-0002-6373-3637}
\email{[firstname.lastname]@unibw.de}
\affiliation{%
  \institution{Universität der Bundeswehr München, RI CODE}
  \city{Neubiberg}
  \country{Germany}
}

\renewcommand{\shortauthors}{Makowski and Pöhn}

\begin{abstract}

Risk-based authentication (RBA) aims to protect end-users against attacks involving stolen or otherwise guessed passwords without requiring a second authentication method all the time. Online services typically set limits on what is still seen as normal and what is not, as well as the actions taken afterward. Consequently, RBA monitors different features, such as geolocation and device during login. If the features' values differ from the expected values, then a second authentication method might be requested. However, only a few online services publish information about how their systems work. This hinders not only RBA research but also its development and adoption in organizations. In order to understand how the RBA systems online services operate, black box testing is applied. To verify the results, we re-evaluate the three large providers: Google, Amazon, and Facebook. Based on our test setup and the test cases, we notice differences in RBA based on account creation at Google. Additionally, several test cases rarely trigger the RBA system. Our results provide new insights into RBA systems and raise several questions for future work.
\end{abstract}



\maketitle

\section{Introduction}
\label{sec:introduction}

Despite the well-established problems of password-based authentication, it continues to be the predominant authentication method used on the Internet~\cite{6234436}. Passwords that are hard to guess by attackers are typically also hard to remember for users~\cite{179476}. Consequently, users often tend to create weaker passwords to avoid the cognitive burden~\cite{5461951}. Although password managers exist, they are often not used efficiently or at all~\cite{281310,238303,263832}. For example, end-users with browser-based password managers tend to create weak passwords and reuse them across accounts for convenience~\cite{281310,238303}. Elderly persons stated fears of single-point-of-failure and the importance of having control over one's data as reasons for not applying password managers~\cite{263832}. In addition, the initial setup of the password manager is cumbersome as the end-user has to add all accounts, preferably changing weak and leaked passwords. This is problematic since online services require user accounts and hence authentication methods, whereas the amount of data and services, such as Internet of Things (IoT) applications, increase~\cite{8436400,icissp23}. Recent data breaches and the following automated credential stuffing attacks further escalate the problem of password-based authentication~\cite{272138,10.1145/3133956.3134067,10.1145/3538969.3544430,208153}.

In order to increase account security, online services often enable two-factor authentication (2FA), where the user has to utilize a second authentication method in addition to the password. As the second factor is always requested, it does not only increase account security but also the burden on users~\cite{8802493,205698}. However, if it is only mandatory for selected systems that deal with sensitive information, the acceptance rate is higher~\cite{10.1145/3313831.3376457}. Nonetheless, second factors can also be bypassed, for example, by phishing~\cite{9945766} and push notifications~\cite{10.1145/3433210.3453084}. Consequently, authentication attempts and all other steps within the authentication lifecycle have to be monitored~\cite{app13042349} to identify attacks, such as what happened at FireEye in the case of SolarWinds' Orion attack~\cite{Yang2022,10.1145/3538969.3544430}. In addition, users might try to avoid 2FA or apply less secure user behavior, such as sharing 2FA tokens due to usability issues, resulting in reduced security~\cite{8998510}.

Risked-based authentication (RBA) in contrast, applies the additional security layer of 2FA only if necessary. RBA estimates whether a login attempt is legitimate or a malicious attempt of account takeover. By monitoring additional features of the login context, such as the Internet protocol (IP) address, IP geolocation, login time, and various fingerprints, RBA systems calculate a risk score related to that login attempt. If these features differ widely from those expected, a second factor might be required to continue. The actions taken depend on the access threshold, which is classified into low, medium, and high according to Wiefling et al.~\cite{10.1007/978-3-030-22312-0_10}. For low risk, typically no actions are needed, whereas for medium or high risk, it might require an additional method of authentication or even block access. The exact access thresholds impact the overall usability and security~\cite{10.1007/978-3-662-64331-0_19}. 
Thereby, RBA strengthens password-based authentication while having fewer usability issues~\cite{10.1145/3427228.3427243} -- at least in theory. Several government agencies, such as the U.S. National Institute of Standards and Technology (NIST)~\cite{nistsp80063b}, recommend or mandate RBA to protect users from password-based attacks. Additionally, online services started to implement RBA. However, only little information about RBA algorithms and configurations is published, see, for example~\cite{206258,openam,8397628}. In order to understand the RBA system of online services, black box testing is applied by Wiefling et al.~\cite{10.1007/978-3-662-64331-0_19,10.1145/3546069}. With the use of Single Sign-On (SSO), the security of the accounts used for authentication becomes more important, as shown by Gavazzi et al.~\cite{sso}. These are typically big services, such as Google, Amazon, Facebook, and Apple.

To verify the results of Wiefling et al.'s study corresponding to RBA, we re-evaluate the three large providers, Google, Amazon, and Facebook, in our study. In our test setup, we apply differently created and configured accounts to evaluate the influence of these parameters. In addition, we include several features of the login context in numerous test cases. We confirm the observation of Wiefling et al.~\cite{10.1145/3546069} that RBA is rarely triggered at several services. Going beyond previous work, the results show that RBA configurations might even be more complex than known, as RBA systems react differently depending on how the account was set up and which fallback methods are configured. Though we could not prove Facebook's VIP status described by Wiefling et al.~\cite{10.1007/978-3-030-22312-0_10}. This might be the result of the comparably short study duration of three months.

The remainder of the paper is as follows: We first give an overview of related work in Section~\ref{sec:sota}. In Section~\ref{sec:methods}, we describe the methodology applied by our study. Next, we outline the results (see Section~\ref{sec:results}) and evaluate the findings (see Section~\ref{sec:evaluation}). This is followed by a discussion of the study and outline of future work in Section~\ref{sec:discussion}. Lastly, we summarize the approach and the results of this paper.

\section{Related Work}
\label{sec:sota}

As outlined beforehand, work related to RBA is difficult to find. Arias-Cabarcos et al.~\cite{10.1145/3336117} provide an overview of RBA by analyzing publications in a structured survey. Bumiller et al.~\cite{10.1145/3582696} present an updated literature review and a classification of existing proposals. Both approaches propose different future work directions.

Several approaches try to shed light on the characteristics of RBA and its configuration. Markert et al.~\cite{281234} analyze how administrators would configure RBA. The authors reason that a default setting is important for orientation. Freeman et al.~\cite{206258} utilize a statistical approach to measure the users' authenticity by applying different features such as IP, geolocation, browser configuration, and time of day. Wiefling et al.~\cite{10.1007/978-3-030-22312-0_10} evaluate the large services of Google, Facebook, and Amazon by black box testing. We integrate and extend their features of the login context and test setup in our study to re-evaluate the results. Wiefling et al.~\cite{10.1145/3427228.3427243} study the usability and security perception of RBA, followed by an analysis of RBA characteristics~\cite{10.1007/978-3-662-64331-0_19}. We incorporate their results. Lastly, Wiefling et al.~\cite{10.1145/3546069} collected feature data from 3.3 million users and 31.3 million login attempts over more than a year. This serves as the basis for information on characteristics, configurations, machine-learning parameters, and a synthesized RBA dataset, among others. Similarly, Papaioannou et al.~\cite{9966901} provide a dataset for behavioral biometrics used on smartphones.

Last but not least, some approaches concentrate on specific aspects of RBA. Wiefling et al. consider the privacy of RBA systems~\cite{9583699} and re-authentication~\cite{10.1007/978-3-030-58201-2_19}. Several authors focus on the evaluation of fingerprinting and behavioral biometrics for RBA~\cite{10.1145/2751323.2751329,10.1145/2695664.2695908,10.1145/3546118.3546152,10.1007/978-3-031-16815-4_30}, whereas Misbahuddin et al.~\cite{8397628} design a risk engine. Rivera et al.~\cite{10.1145/3411508.3421377} study the detection of user authentication attacks involving network tunneling geolocation deception, Papaioannou et al.~\cite{9966915} estimate the risk in the use case of mobile passenger ID devices, and Wiefling et al.~\cite{10.1007/978-3-030-35055-0_12} mimic humanoid usage behavior for future studies. 

To conclude, although some approaches and studies have been published, almost no data about the applied RBA algorithms and features is available. As the last black box test was conducted in 2019, we re-evaluate the findings by integrating the results of the related work and additionally taking different account creations and configurations into account.

\section{Methodology}
\label{sec:methods}

In this section, we outline our methodology, including account creation and tested features. Additionally, we consider ethics and limitations. The study was mainly conducted in November and December 2022, with preparations beforehand and analysis afterward.

\subsection{Account Creation}

Typically, account creation should be as convenient as possible. At the same time, website providers are interested in keeping the number of fake accounts to a minimum to prevent spam or illegitimate usage. Different approaches are applied to combat this problem, ranging from phone numbers tied to a maximum number of accounts to freezing accounts after creation if unwanted behavior is detected. Consequently, we had to make sure that our accounts did not break any unwritten rules. On the other hand, different account settings were taken into account to evaluate if they resulted in the same behaviors. To make the accounts as realistic as possible without linking them to real people, we created fake identities with names, birthdates, addresses, and artificial intelligence (AI)-generated profile pictures. In addition, we added different fallback authentication methods, ranging from recovery email and phone number to logging in to the app. The accounts were created in a timespan of one month. We noticed the following differences and limitations when creating several accounts.

\begin{description}
\item[Google:] Requires a phone number per newly created account, while a phone number can be used for two accounts. An exception applies for Android phones if the account is opened through the settings app. In this case, no phone number is necessary. We are unsure whether Google tracks the number if a Subscriber Identification Module (SIM) card is inserted.
\item[Amazon:] Requires a unique phone number for each. Removing a phone number from an account does not result in the possibility of using it on at another account. 
\item[Facebook:] Requires a phone number \emph{or} email address. When trying to create multiple accounts in quick succession, Facebook immediately blocks any other accounts after they are created.
\end{description}

We set up the following accounts for Google (see Table~\ref{tab:gtests}) and Amazon (see Table~\ref{tab:atests}). To test Facebook's VIP status as stated by Wiefling et al.~\cite{10.1007/978-3-030-22312-0_10}, we set up six different account characteristics shown in Table~\ref{tab:fbtests}.

\begin{table}[!htpb]
\centering
	\caption{Setup for Google tests}
	\label{tab:gtests}
	\begin{tabular}{ll}\toprule
		 \textbf{Name}       & \textbf{Description}                                                \\ \midrule
		G1             & Desktop account, phone number + email                                    \\
		G2          & Desktop account, phone number                                       \\
		G3           & Mobile account, no recovery information                                    \\
		G4             & Mobile account, phone number + email                                      \\
		G5             & Mobile account, phone number                                        \\ \bottomrule
	\end{tabular}
\end{table}

\begin{table}[!htpb]
\centering
	\caption{Setup for Amazon tests}
	\label{tab:atests}
	\begin{tabular}{ll}\toprule
		 \textbf{Name}       & \textbf{Description}                                                \\ \midrule
		A1             & Email, phone number removed                                   \\
		A2          & Phone number                                      \\
		A3           & Phone number + app                                    \\  \bottomrule
	\end{tabular}
\end{table}

\begin{table}[!htpb]
\centering
	\caption{Setup for Facebook VIP tests}
	\label{tab:fbtests}
	\begin{tabular}{ll}\toprule
		 \textbf{Name}       & \textbf{Description}                                                \\ \midrule
		FB1             & New account without any activity                                    \\
		FB2          & New account with five friends                                       \\
		FB3           & New account, joined five groups                                     \\
		FB4             & New account, five posts liked                                       \\
		FB5             & New account, five posts made                                        \\
		FB6            & Account 1 month old + FB2-4                 \\\bottomrule
	\end{tabular}
\end{table}

\subsection{Tested Features}

This section explains the different types of tests, their intentions, and the necessary preparations.

\subsubsection{Geolocation}

A significant factor in RBA analysis is location data. Because IP address and IP geolocation are typically linked, both factors are tested together to understand their impact. Table~\ref{tab:geotests} lists all tests and gives a brief description. The test cases are ordered by distance from the known location \emph{home} with the same Internet service provider (ISP). To increase confidence, the tests are repeated on different days and on multiple accounts. Test cases IP6 to IP9 explore the effects of crossing borders and differentiating between neighboring states and continents. To keep results consistent and enable easy repeatability, every test abroad is conducted with the help of a virtual private network (VPN).
Lastly, IP10 examines if the TOR network has visible effects on the risk score.

\begin{table}[!htpb]
	\centering
	\caption{Setup for geolocation tests}\label{tab:geotests}
	\begin{tabular}{lp{6.3cm}}
		\toprule
		     \textbf{Name}               & \textbf{Description}                                                         \\ \midrule
		IP1          & Known location and ISP                                                       \\
		IP2             & Known location with known SIM                                           \\
		IP3          & Wi-Fi in an unknown location                                                 \\
		IP4             & Cellular in an unknown location                                              \\
		IP5    & Different city, unknown location, travel time taken into account             \\
		IP6           & VPN connection to a neighboring state, travel time taken into account   \\
		IP7         & VPN connection to a neighboring state, instant travel time              \\
		IP8    & VPN connection to a different continent, travel time taken into account \\
		IP9  & VPN connection to a different continent, instant travel time            \\
		IP10                  & Using known TOR exit nodes                                                   \\ \bottomrule
	\end{tabular}
\end{table}

\subsubsection{Device}

Features such as language, resolution, and installed apps make a device unique and influence the risk score. Generally, every new or unknown device should raise a RBA reaction. In Table~\ref{tab:devtests}, D2 tests the behavior when utilizing anonymous browsing tools supplied with the browser. Tests D3-D5 use \emph{``soft factors''}, whereas resolution in D6 generally rarely changes. To alter the user agent string (UAS) in D7, the extension User-Agent Switcher for Chrome~\cite{uas} is utilized, verified with Cover Your Tracks~\cite{eff}. D8 and D9 apply the Sensors tab in the Google Chrome developer tools~\cite{dev} to change only the location or location and resolution together.

\begin{table}[!htpb]
	\centering
	\caption{Setup for device tests}\label{tab:devtests}
	\begin{tabular}{ll}
		\toprule
		   \textbf{Name}          & \textbf{Description}                                    \\ \midrule
		D1           & Known device                                            \\
		D2       & Anonymous browsing tools, e.\,g., incognito mode  \\
		D3         & Different language                                        \\
		D4               & Different login time                                      \\
		D5             & Different time zone setting                               \\
		D6          & Different screen resolution                             \\
		D7         & Different UAS                                      \\
		D8         & Chrome's locations changing tools                       \\
		D9  & Locations changing tools and resolution change \\ \bottomrule
	\end{tabular}
\end{table}

\subsubsection{SIM Card}

The SIM card grants access to a mobile carrier's network, facilitates billing, and allows usage monitoring. It also associates the phone with a phone number and enables calls and Internet usage. For these reasons, a SIM card must be unique. The tests in Table~\ref{tab:simtest} aim to determine if services have access to and use SIM card data and whether the presence of a SIM card changes RBA behavior. This study utilizes SIM cards from different providers.

\begin{table}[]
	\centering
	\caption{Setup for SIM card tests}\label{tab:simtest}
	\begin{tabular}{ll}
		\toprule
		    \textbf{Name}      & \textbf{Description}                      \\ \midrule
		S1        & Known device with no SIM card        \\
		S2 & Known device with unknown SIM card   \\
		S3    & Different device with known SIM card \\ \bottomrule
	\end{tabular}
\end{table}

\subsection{Ethical Considerations and Limitations}

The study was in compliance with the university review board and, consequently, did not require specific approval. The accounts applied artificial data (e.\,g., name and picture) to reduce the connection to real persons. Nonetheless, this can have effects on the observed behavior, such as friends not being real friends on Facebook. Due to the short duration of the study, the number and age of accounts are limited. Especially, age can result in not gaining Facebook's VIP status. Additional effects might be observable over a longer study period, such as learning effects. Although we tried to separate all accounts from each other by SIM cards, account data, and devices, indirect factors such as the IP address or the login time could link them.

\section{Results}
\label{sec:results}

In this section, we present our results of the geolocation, device, and SIM card tests.

\subsection{Geolocation}

The geolocation tests were conducted by actually changing locations, using a VPN or a TOR browser, depending on the required test and the accuracy of the location. 

\begin{table*}[!htbp]
	\caption[Geo test results]{Geo test results}\label{tab:GeoResults}
    \begin{subtable}[c]{0.30\textwidth}
		\centering
		\caption{Google}\label{tab:googleGeoResults}
		
		\begin{tabular}{@{}lccccc@{}}
		\toprule
			\rotatebox{\tableturnl}{test}  & \rotatebox{\tableturnl}{G1 } & \rotatebox{\tableturnl}{G2 } & \rotatebox{\tableturnl}{G3 } & \rotatebox{\tableturnl}{G4 } & \rotatebox{\tableturnl}{G5 } \\ \midrule
			IP1  &  \gball   &  \gball   &  \gball   &  \gball   &  \gball   \\ 
			IP2  &  \gball   &  \gball   &  \gball   &  \gball   &  \gball   \\ 
			IP3  &  \oball   &  \oball   &  \gball   &  \gball   &  \gball   \\ 
			IP4  &  \gball   &  \gball   &  \gball   &  \gball   &  \gball   \\ 
			IP5  &  \gball   &  \gball   &  \gball   &  \gball   &  \gball   \\ 
			IP6  &  \oball   &  \oball   &  \oball   &  \oball   &  \gball   \\ 
			IP7  &  \oball   &  \oball   &  \gball   &  \gball   &  \gball   \\ 
			IP8  &  \oball   &  \oball   &  \oball   &  \oball   &  \oball   \\ 
			IP9  &  \oball   &  \oball   &  \gball   &  \gball   &  \gball   \\ 
			IP10 &  \oball   &  \oball   &  \oball   &  \gball   &  \oball   \\ 
			\bottomrule
		\end{tabular}
    \end{subtable}
    \begin{subtable}[c]{0.30\textwidth}
		\centering
		\caption{Amazon}\label{tab:amazonGeoResults}
		\begin{tabular}{@{}lccc@{}}
		\toprule
			\rotatebox{\tableturnl}{test}  & \rotatebox{\tableturnl}{A1 }& \rotatebox{\tableturnl}{A2 }& \rotatebox{\tableturnl}{A3 } \\ \midrule
			IP1  &  \gball   &  \gball   &  \gball   \\ 
			IP2  &  \gball   &  \gball   &  \gball   \\ 
			IP3  &  \gball   &  \gball   &  \gball   \\ 
			IP4  &  \gball   &  \gball   &  \gball   \\ 
			IP5  &  \gball   &  \gball   &  \gball   \\ 
			IP6  &  \oball   &  \gball   &  \oball   \\ 
			IP7  &  \gball   &  \oball   &  \gball   \\ 
			IP8  &  \gball   &  \gball   &  \gball   \\ 
			IP9  &  \gball   &  \gball   &  \gball   \\ 
			IP10 &  \gball   &  \gball   &  \gball   \\ 
			\bottomrule
		\end{tabular}

    \end{subtable}
    \begin{subtable}[c]{0.30\textwidth}
\centering
	\caption{Facebook}\label{tab:facebookGeoResults}
	\begin{tabular}{@{}lcccccc@{}}
	\toprule
		\rotatebox{\tableturnl}{test} &  \rotatebox{\tableturnl}{FB1 }   &  \rotatebox{\tableturnl}{FB2 }   &  \rotatebox{\tableturnl}{FB3 }   &  \rotatebox{\tableturnl}{FB4 }   &  \rotatebox{\tableturnl}{FB5 }   &  \rotatebox{\tableturnl}{FB6 }   \\ \midrule
		IP1  & \gball & \gball & \gball & \gball & \gball & \gball \\ 
		IP2  & \gball & \bball & \gball & \gball & \bball & \gball \\ 
		IP3  & \gball & \gball & \gball & \gball & \gball & \gball \\ 
		IP4  & \gball & \gball & \gball & \gball & \gball & \gball \\ 
		IP5  & \gball & \bball & \gball & \gball & \gball & \bball \\ 
		IP6  & \rball & \bball & \gball & \gball & \rball & \bball \\ 
		IP7  & \rball & \rball & \rball & \rball & \rball & \gball \\ 
		IP8  & \rball & \bball & \gball & \rball & \rball & \bball \\ 
		IP9  & \rball & \bball & \rball & \rball & \rball & \bball \\ 
		IP10 & \rball & \bball & \rball & \rball & \rball & \bball \\ 
		\bottomrule
	\end{tabular}
    \end{subtable}
	\parbox[][3em][c]{\linewidth}{\centering
		\gball~: successful login\quad
\oball~: RBA triggered\quad
		\rball~: RBA triggered, account locked out\quad
		\bball~: not tested, account was blocked}
\end{table*}

\begin{figure*}[!htpb]
	\centering
\begin{subfigure}[b]{0.3\textwidth}
	\includegraphics[width=\textwidth]{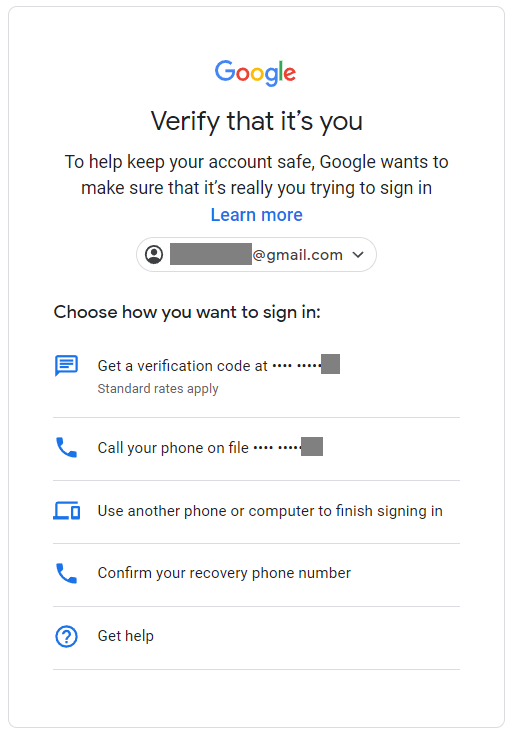}
	\caption{Account with phone number}\label{fig:phone}
\end{subfigure}
	\hfill
\begin{subfigure}[b]{0.3\textwidth}
	\includegraphics[width=\textwidth]{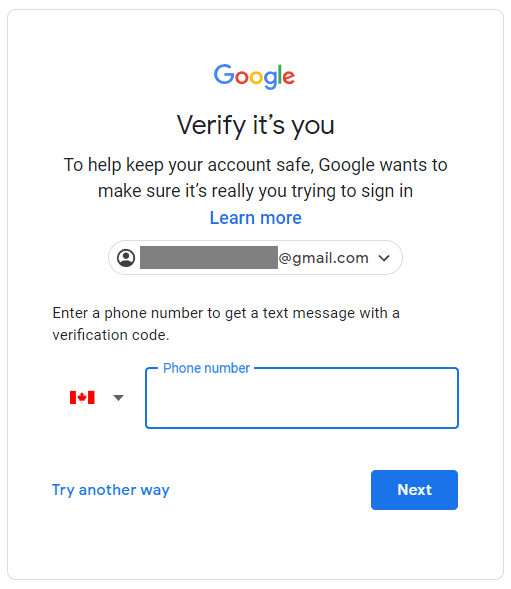}
	\caption{Account without any recovery method}\label{fig:number}
\end{subfigure}
	\hfill
\begin{subfigure}[b]{0.3\textwidth}
	\includegraphics[width=\textwidth]{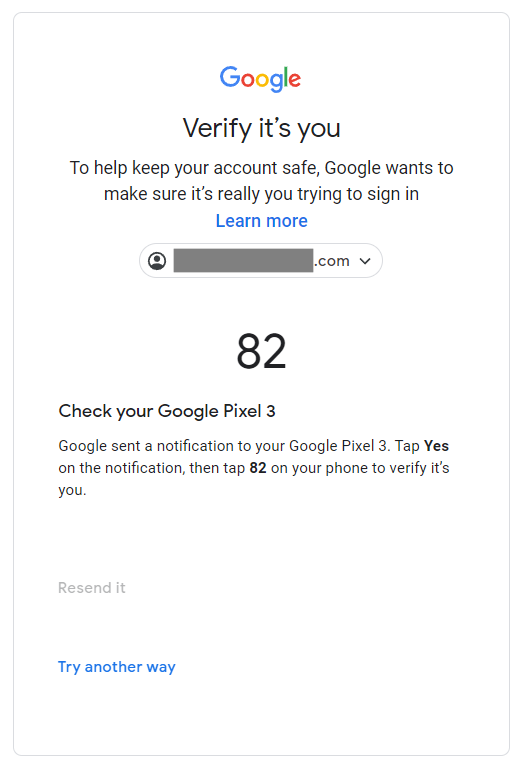}
	\caption{Account signed in on Android phone}
\end{subfigure}
	\caption{Google's different RBA screens}
	\label{fig:googleRBAscreens}
	\Description[Google's different RBA screens]{Google's different RBA screens: account with phone number behaves differently than an account without any recovery method or an account signed in on Android phone}
\end{figure*}

Table~\ref{tab:GeoResults} gives an overview of all geolocation test results. On the left, in Table~\ref{tab:googleGeoResults}, the results of Google are given. As one would expect, IP1 with no change at all showed no unusual behavior. The first finding occurs on IP3 (public Wi-Fi), where Google's RBA system triggered, and the user was forced to complete one common RBA task. Interestingly, only those accounts that were created with a phone number provoked an RBA response; the other accounts, even if a phone number was added later, did not. This scheme continued through most of the tests. IP4 was conducted shortly after IP3 and within the same vicinity; this could explain why all log-in attempts were granted immediately. The tests with unrealistic travel times (IP7 and IP9) and the TOR test case (IP10) triggered the RBA system. This is in line with the findings of Wiefling et al.~\cite{10.1007/978-3-030-22312-0_10}. However, just like the tests before, only the accounts with mandatory phone numbers had an RBA reaction, but tests IP6 and IP8 (realistic travel time) triggered the system on most accounts. The TOR test (IP10) causes an additional Completely Automated Public Turing test to tell Computers and Humans Apart (CAPTCHA) after the email address was input. Interestingly, IP8, which was conducted after two days of no login, was the first test where all accounts provoked an RBA response in various forms. The differing account setups revealed themselves on the different RBA screens. Figure~\ref{fig:googleRBAscreens} shows the other methods to log in after the system was triggered. If an account cannot receive a login code, the screen displayed in Figure~\ref{fig:number} requests a phone number. Curiously, after providing a number and logging in with the code received, the number is not added to the account. Sometimes, Google would suggest a code via short message service (SMS) or phone call, but other methods would still be accessible.

Facebook did not show any behavior at IP1 (baseline) or IP3-IP5. However, we did accidentally find out that Facebook would allow typos in the email address during login. The test cases concerning changes in the IP address triggered the RBA system, blocking access to the account without any information about the reasons. Similarly to Google, the system reacted the same across all foreign tests. After a few attempts, Facebook blocked two of the accounts (including FB6), shown with gray circles in Table~\ref{tab:facebookGeoResults}. Although uploading a photo of the user lifted the suspension, it got blocked again after further logins. The results for IP6 were mixed: half had no reaction and half had been blocked. Since the blocked accounts included low and high activity profiles, no meaningful assumption about the VIP state can be made.

Amazon's RBA system seems to be the weakest of all tested, as displayed in Table~\ref{tab:amazonGeoResults}. Requiring the user to retype the password in addition to solving a CAPTCHA does not deter an attacker. The RBA prompts were inconsistent across accounts, and it seemed that solving a CAPTCHA once prevented it from appearing the next few times. Apart from notifications via SMS or email, the site itself never reacted to failed login attempts. Curiously, the repeated logins in quick succession triggered some form of rate limiting multiple times. Either the website would be a largely simplified version or an alert about being too fast would prevent the user from continuing.

\subsection{Devices}

\begin{table*}[!htbp]
		\caption{Device test results}\label{tab:DeviceResults}
    \begin{subtable}[c]{0.30\textwidth}
		\centering
		\caption{Google}\label{tab:googleDeviceResults}
		\begin{tabular}{@{}lccccc@{}}
			\toprule
			\rotatebox{\tableturnl}{test}  & \rotatebox{\tableturnl}{G1 } & \rotatebox{\tableturnl}{G2 } & \rotatebox{\tableturnl}{G3 } & \rotatebox{\tableturnl}{G4 } & \rotatebox{\tableturnl}{G5 } \\ \midrule
			D1   &  \gball   &  \gball   &  \gball   &  \gball   &  \gball   \\ 
			D2   &  \rball   &  \rball   &  \gball   &  \gball   &  \gball   \\ 
			D3   &  \gball   &  \gball   &  \gball   &  \gball   &  \gball   \\ 
			D4   &  \gball   &  \gball   &  \gball   &  \gball   &  \gball   \\ 
			D5   &  \gball   &  \gball   &  \gball   &  \gball   &  \gball   \\ 
			D6   &  \gball   &  \gball   &  \gball   &  \gball   &  \gball   \\ 
			D7   &  \gball   &  \gball   &  \gball   &  \gball   &  \gball   \\ 
			D8   &  \gball   &  \gball   &  \rball   &  \gball   &  \gball   \\ 
			D9   &  \gball   &  \gball   &  \gball   &  \gball   &  \gball   \\  \bottomrule
		\end{tabular}
    \end{subtable}
    \begin{subtable}[c]{0.30\textwidth}
		\centering
		\caption{Amazon}\label{tab:amazonDeviceResults}
		\begin{tabular}{@{}lccc@{}}
			\toprule
			\rotatebox{\tableturnl}{test}  & \rotatebox{\tableturnl}{A1 }& \rotatebox{\tableturnl}{A2 }& \rotatebox{\tableturnl}{A3 } \\ \midrule
			D1   &  \gball   &  \gball   &  \gball   \\ 
			D2   &  \gball   &  \gball   &  \gball   \\ 
			D3   &  \gball   &  \gball   &  \gball   \\ 
			D4   &  \gball   &  \gball   &  \gball   \\ 
			D5   &  \gball   &  \oball   &  \oball   \\ 
			D6   &  \oball   &  \oball   &  \oball   \\ 
			D7   &  \gball   &  \oball   &  \gball   \\ 
			D8   &  \gball   &  \gball   &  \gball   \\ 
			D9   &  \gball   &  \gball   &  \gball   \\  \bottomrule
		\end{tabular}
		
	
    \end{subtable}
    \begin{subtable}[c]{0.30\textwidth}
\centering
	\caption{Facebook}\label{tab:facebookDeviceResults}
	\begin{tabular}{@{}lcccccc@{}}
		\toprule
		\rotatebox{\tableturnl}{test} &  \rotatebox{\tableturnl}{FB1 }   &  \rotatebox{\tableturnl}{FB2 }   &  \rotatebox{\tableturnl}{FB3 }   &  \rotatebox{\tableturnl}{FB4 }   &  \rotatebox{\tableturnl}{FB5 }   &  \rotatebox{\tableturnl}{FB6 }   \\ \midrule
		D1   & \gball & \bball & \gball & \gball & \gball & \bball \\ 
		D2   & \gball & \bball & \gball & \gball & \gball & \bball \\ 
		D3   & \gball & \bball & \gball & \gball & \gball & \bball \\ 
		D4   & \gball & \bball & \gball & \gball & \gball & \bball \\ 
		D5   & \gball & \bball & \gball & \gball & \gball & \bball \\ 
		D6   & \gball & \bball & \gball & \gball & \gball & \bball \\ 
		D7   & \gball & \bball & \gball & \gball & \gball & \bball \\ 
		D8   & \gball & \bball & \gball & \gball & \gball & \bball \\ 
		D9   & \gball & \bball & \gball & \gball & \gball & \bball \\  \bottomrule
	\end{tabular}
    \end{subtable}
	\parbox[][3em][c]{\linewidth}{\centering
	\gball~: successful login\quad
\oball~: RBA triggered\quad
\rball~: RBA triggered, password reset\quad
	\bball~: not tested, account was blocked}
\end{table*}

These tests employed Google Chrome's incognito mode and developer tools in conjunction with a UAS changer add-on. Crucially, the location was set with the developer tools exclusively, never with a VPN. During the tests D1-D7, Google only had one interesting reaction: Both accounts with mandatory phone numbers were forced to reset their password, which means a risk score between \emph{medium} and \emph{high} based on Wiefling et al. was triggered. As users were not locked out of their accounts entirely, setting a new password was enough to continue browsing, see Table~\ref{tab:googleDeviceResults}. Google's RBA system did not react to the other accounts, which were made on an Android phone without a phone number. Additionally, the rest of the tests did not trigger the system, except for one case in D8, where one account required a new password.

Since the device tests were conducted chronologically after the geolocation tests, two of the Facebook accounts were blocked and not available for testing. Nevertheless, all other accounts managed to log in without problems or RBA prompts. Similarly to Google, Facebook changes the site's language depending on browser preferences. Table~\ref{tab:facebookDeviceResults} summarizes the results.

Amazon's behavior was more nuanced than Facebook's, as displayed in Table~\ref{tab:amazonDeviceResults}. D1-D4 showed no change in the login procedure. D5 (unusual login time zone) prompted to retype the password together with a CAPTCHA. Because no further actions were taken, we assume that this was a security measure against automated password attacks. The website did not change the language and even informed the user that only goods available for shipping to Germany were displayed, presumably because Amazon knew the location from the IP address.

\subsection{SIM Cards}

\begin{table*}[!htbp]
		\caption{SIM test results}\label{tab:SimResults}
    \begin{subtable}[c]{0.30\textwidth}
		\centering
		\caption{Google}\label{tab:googleSimResults}
		\begin{tabular}{@{}lccccc@{}}
			\toprule
			\rotatebox{\tableturnl}{test}  & \rotatebox{\tableturnl}{G1 } & \rotatebox{\tableturnl}{G2 } & \rotatebox{\tableturnl}{G3 } & \rotatebox{\tableturnl}{G4 } & \rotatebox{\tableturnl}{G5 } \\ \midrule
			S1   &  \gball   &  \gball   &  \gball   &  \gball   &  \gball   \\ 
			S2   &  \oball   &  \oball   &  \rball   &  \oball   &  \oball   \\ 
			S3   &  \oball   &  \oball   &  \rball   &  \oball   &  \oball   \\  \bottomrule
		\end{tabular}
    \end{subtable}
    \begin{subtable}[c]{0.30\textwidth}
		\centering
		\caption{Amazon}\label{tab:AmazonSimResults}
		\begin{tabular}{@{}lccc@{}}
			\toprule
			\rotatebox{\tableturnl}{test}  & \rotatebox{\tableturnl}{A1 }& \rotatebox{\tableturnl}{A2 }& \rotatebox{\tableturnl}{A3 } \\ \midrule
			S1   & \gball    & \gball    & \gball        \\ 
			S2   & \gball    & \gball    & \gball        \\ 
			S3   & \gball    & \gball    & \gball        \\  \bottomrule
		\end{tabular}
		
    \end{subtable}
    \begin{subtable}[c]{0.30\textwidth}
	\caption{Facebook}\label{tab:facebookSimResults}
	\begin{tabular}{@{}lcccccc@{}}
		\toprule
		\rotatebox{\tableturnl}{test} &  \rotatebox{\tableturnl}{FB1 }   &  \rotatebox{\tableturnl}{FB2 }   &  \rotatebox{\tableturnl}{FB3 }   &  \rotatebox{\tableturnl}{FB4 }   &  \rotatebox{\tableturnl}{FB5 }   &  \rotatebox{\tableturnl}{FB6 }   \\ \midrule
		S1   & \gball & \bball & \gball & \gball & \gball & \bball \\ 
		S2   & \gball & \bball & \gball & \gball & \gball & \bball \\ 
		S3   & \gball & \bball & \gball & \gball & \gball & \bball \\ \bottomrule
	\end{tabular}
    \end{subtable}
	\parbox[][3em][c]{\linewidth}{\centering\gball~: successful login\quad \oball~: RBA triggered\quad \rball~: RBA triggered, login blocked \quad\bball~: not tested, account was blocked}
\end{table*}

The goal of the SIM card tests was to determine whether the websites would react to a change in SIM card or device. While services should not be able to access SIM card data such as phone numbers directly, the mobile connection would reflect the provider. Table~\ref{tab:googleSimResults} shows Google's behavior towards SIM card changes. Unsurprisingly, removing the card did not have an effect on the login, as the device was familiar to the system. However, a SIM card from a different provider provoked an RBA response. All accounts with RBA information attached, e.\,g., phone number or email address, were able to log in after passing the prompt. We did not observe any difference between desktop and phone-created accounts. Account 3 (without any information) was not able to log in because Google had no way of verifying the user's identity.

\begin{figure*}[!htpb]
	\centering
\begin{subfigure}{0.45\linewidth}
\centering
\includegraphics[width=0.8\linewidth]{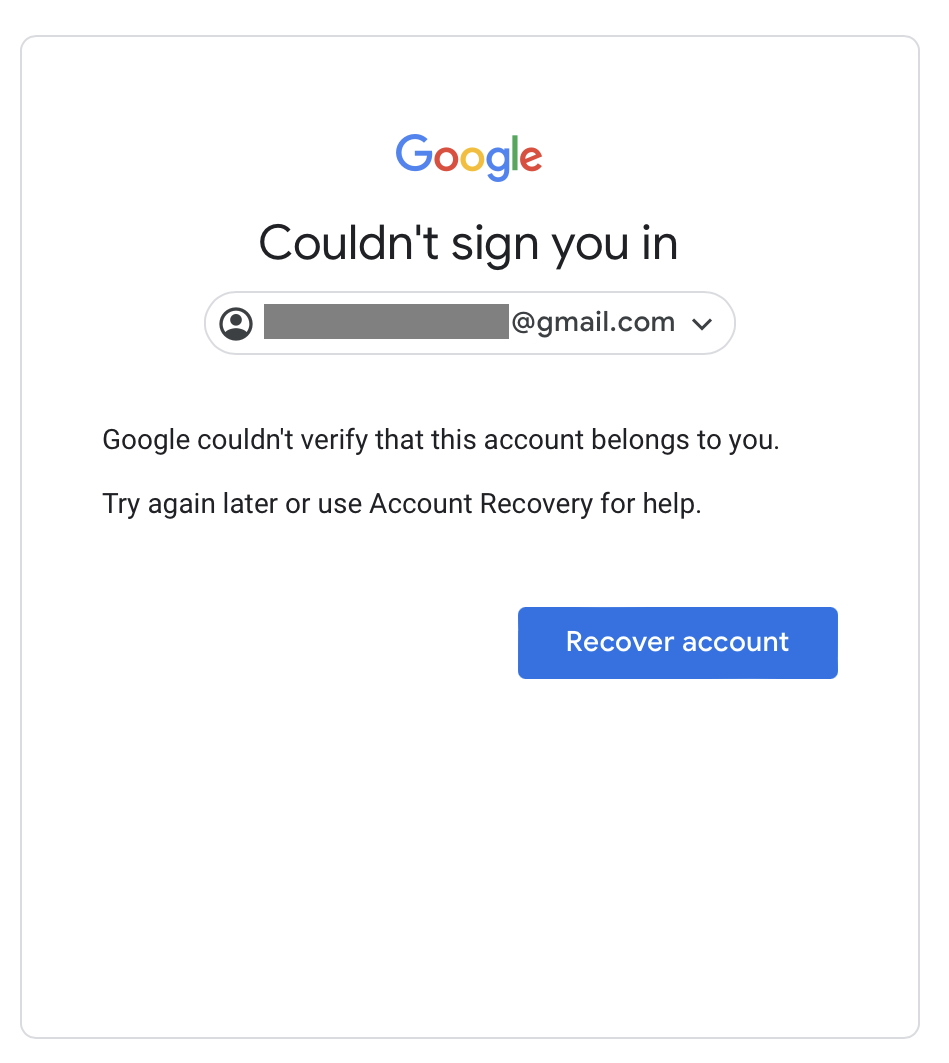}
	\caption{First prompt after login}
\label{fig:firstprompt}
\end{subfigure}
	\hfill
\begin{subfigure}{0.45\linewidth}
\centering
	\includegraphics[width=0.8\linewidth]{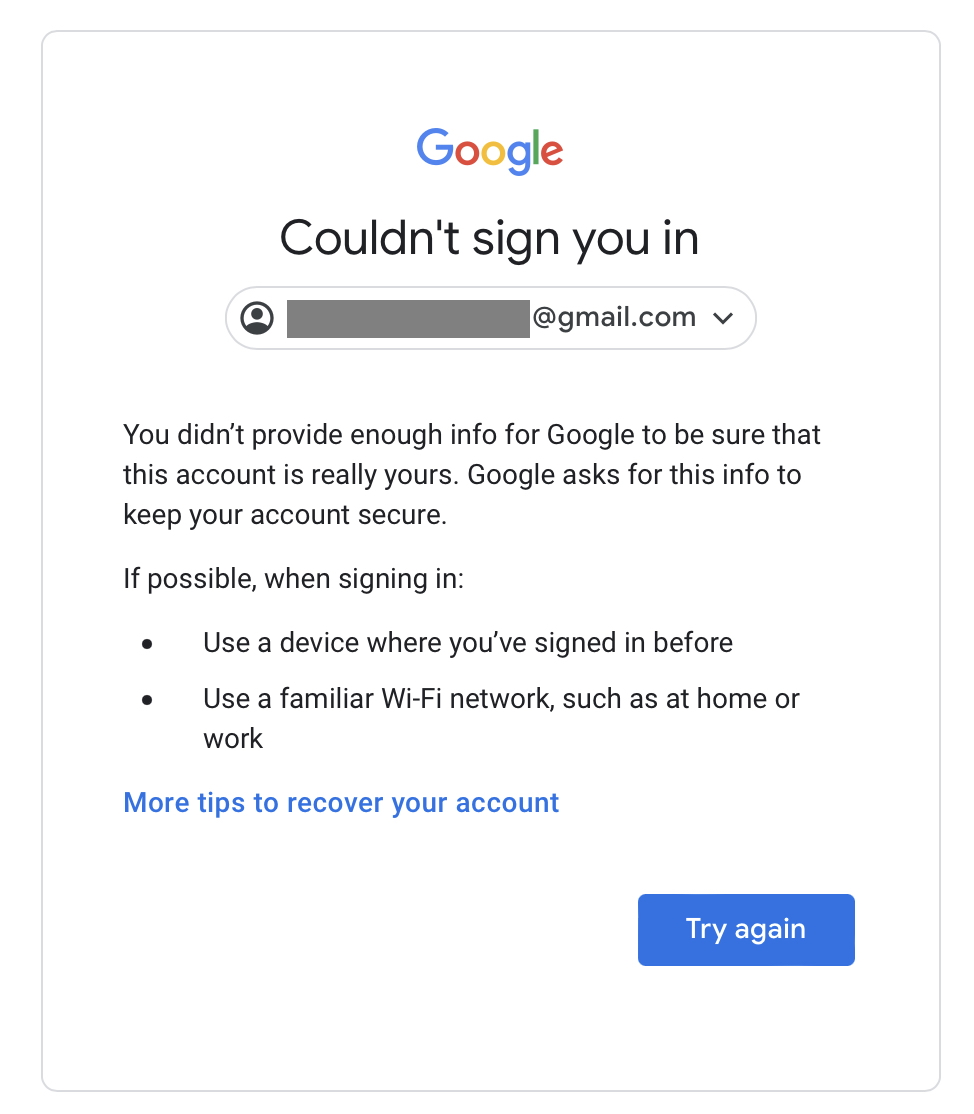}
	\caption{Second prompt after clicking on ``Recover account''}
\label{fig:secondprompt}
\end{subfigure}
	\caption{Google blocks account without RBA information}\label{fig:googlerbablock}
	\Description[Google blocks account without RBA information]{Google blocks account without RBA information}
\end{figure*}

Figure~\ref{fig:googlerbablock} shows the RBA screens, explaining that the user did not provide enough information. Interestingly, during the geolocation tests, the same account could log in using a phone number without it being related to the account. During S3 (different device), Google's system reacted as in S2.
 
Facebook's SIM results show successful logins across all accounts (see Table~\ref{tab:facebookSimResults}). With regard to the results of the other tests, this is expected. Amazon's results are similar to Facebook's, as summarized in Table~\ref{tab:AmazonSimResults}. The absence or change in the SIM card had no effect on the login, and neither did a different device. This is also in line with the other test results.

\section{Evaluation}
\label{sec:evaluation}

More significant changes in IP addresses provoke reactions from Google's RBA system. Google's system adapts to different account setups and provides some form of RBA for all. However, there is an observable difference between accounts created on a desktop computer and ones created on a mobile phone. While this should not have an impact on RBA performance, especially if a number was added after the fact, the tests show a noticeable link between the two types. Subsequent random tests confirmed the results and the distinct difference between realistic and unrealistic travel time tests. The phone accounts only reacted to realistic differences. During the device tests, three of the accounts were prompted to reset their passwords. This happened noticeably earlier for the desktop accounts. Whether this was the product of repeated quick logins from different IP addresses or devices, or the actual RBA system's reaction to this single login, it could not be determined. The results of the SIM card test are as expected. Interestingly, the account with no RBA information was not able to log in. This behavior is different from the other tests, where a phone number was able to resolve the prompt. Whether a random phone number is secure enough to prove the user's identity is questionable. Furthermore, Google alerted the user to this activity right away after a successful login.

The RBA system by Facebook was objectively black and white. During our study, we never saw an ordinary RBA screen. The site either granted access or denied it completely. While allowing logins up to a neighboring state, the site would block every attempt from outside the continent. The device and SIM card tests never provoked the RBA system to react. Facebook's communication about the reasons why the login failed was non-transparent: the site always showed a message about incorrect credentials. The actual user could be left retyping the password several times. Even after a successful login, Facebook does not explain why previous attempts failed or if there has been suspicious activity. Since multiple accounts were blocked right after creation, Facebook could have somehow linked the accounts and flagged each one even before any tests began. Later and repeated blocks of the account support this theory. Lastly, there was no noticeable difference between accounts, and the suspected VIP model could not be observed.

Amazon's RBA system appears to be the weakest of all the ones tested. In addition to being triggered the least amount of times, the RBA screens (retyping the password and a CAPTCHA) were not really protecting the accounts. The RBA prompts were inconsistent across accounts. It seemed that solving a CAPTCHA once prevented it from appearing in the future. Just like Facebook, Amazon did not respond to SIM card changes. Apart from notifications via SMS or email, the site itself never reacted to failed login attempts. Curiously, the repeated logins in quick succession triggered some form of rate limiting (simplified version or alert) multiple times.

\section{Discussion}
\label{sec:discussion}

Although our study is limited due to its duration and inclusion of websites, it partly verifies previous research and provides new insights. The key takeaways of Wiefling et al.~\cite{10.1007/978-3-030-22312-0_10} of when and how a particular RBA system reacts were partly verified. Where Google only sent a security mail in the study, the RBA system already triggered a heightened response. Our study observed similar ways to resolve a blocked login attempt and found two additional paths for accounts with a logged-in phone and accounts with no RBA information at all. A new study could explore the account security features depending on what platform the accounts and maybe the operating system (OS) were created on. The idea of Facebook VIP accounts could not be validated, and few to no changes in behavior were found regarding profile activity. The results of Amazon's tests were not as clear as in related work; after triggering the system once, it generally did not react on later tests. Our results suggest that either there are more than three risk scores or there is more nuanced behavior within the risk scores.

Based on the preliminary results, a study of various accounts per website with different settings could provide insights into the influence of the setting and the device the account was created with. The test cases outlined in this paper can be used for the study. The results could be compared with the configuration of administrators, similar to Markert et al.~\cite{281234}.

We wonder what other factors may play a role in how RBA systems react and if international websites use it for all services or per country or region. A long-term study could reveal further features. An elaborated RBA system should flag a new location as suspicious, potentially distinguishing between faraway locations and close ones. However, it should be able to learn about recent locations, for example, a new workplace, and only ask the user a few times for further authentication. If a system is incapable of learning, it would treat every login attempt like a 2FA system and lose all benefits. Especially, login history size and frequency-dependent re-authentication have not been explored yet. This could be a follow-up study, which would require long-term data collection.

Finally, only the three websites of Google, Facebook, and Amazon were analyzed in related work. However, we assume that more websites apply RBA. For future research, the knowledge of further providers would enhance the results. So far, it is unclear how to get to know that a provider is utilizing RBA. Hence, various ways to detect RBA are regarded. The personal experience of authors and other users could play a role. Nonetheless, further ways should be considered. For example, RBA might be detectable via information on the websites themselves, by analyzing data privacy statements, and data exports according to General Data Protection Regulation (GDPR).

\section{Conclusion}
\label{sec:conclusion}

As the Internet expands, data collection and account creation increase. To keep end-user data secure, online services must protect access to these accounts sufficiently while maintaining usability. One way to enable both is through the application of RBA. Since little information is available on how these systems work, we set up a study with several test cases to evaluate RBA at Google, Facebook, and Amazon. We conducted the study and evaluated the results. We noticed that the account creation and information provided by the user can play a role in the RBA system's reaction. In addition, the RBA systems might be more fine-grained than just low, medium, and high. To evaluate Facebook's VIP status, a longer-term study might be required. Website providers like Google and Facebook try to minimize fake accounts. Since this was not the main focus, the systems were not further investigated. Nevertheless, this might be done in future work. In addition, we want to explore more features, such as existing data, learning effects, and differences in OSs.

\bibliographystyle{ACM-Reference-Format}
\bibliography{rba}

\end{document}